\documentstyle[12pt]{article}
\begin{document}
\title{ n-particle sector of field theory as a quantum open system.}
\author {C. Anastopoulos \\ Theoretical Physics Group, The Blackett Lab. \\
Imperial College \\ E-mail : can@tp.ph.ic.ac.uk \\}
\date { October 1996}
\maketitle
\begin{abstract}
We give an exposition of a technique, based on the Zwanzig projection
formalism, to construct the evolution equation for the reduced density
matrix corresponding to the n-particle sector of a field theory. We
consider the case of a scalar field with a $g \phi^3$ interaction as
an example and construct the master equation at the lowest non-zero
order in perturbation theory.
\end{abstract}

\renewcommand {\theequation}{\thesection.\arabic{equation}}
\let \ssection = \section
\renewcommand{\section}{\setcounter{equation}{0} \ssection}

\section{Motivation}
\par
The Hilbert space of a quantum field theory is the Fock space ${\cal
F}$ constructed from the  n-particle subspaces $H^n$ as ${\cal F} = 
\bigoplus_{n=0}^{\infty} H^n$. It is a well known feature that the
interaction terms cause mixing of the states between these subspaces. 
\par
It is the aim of this letter to offer a treatment of the subspaces
$H^n$ as corresponding to an open quantum system, that is we want to
obtain an evolution equation for the reduced density matrix
corresponding to the n-particle subspace, while treating the rest of
the Fock space as an ``environment''. Our master equation is expected
to contain (in general non-local in time) terms describing dissipation
and noise. In addition, the fact that at any time the system might
leave the relevant subspace the density matrix will not be normalized
to one. The quantity $ 1 - Tr \rho_{rel}(t)$ will correspond to the
probability that at time $t$ the n-particle subspace is ``empty''.
\par
This is a  problem of interest mainly for two reasons.  First, in an
attempt to explain the manifestation of particles as spatially
localized excitations it is important to know the degree of validity
of the particle description.  Our approach of treating the n-particle
sector as an open system provides measures of the modification in the
equations of motion (dissipation) as well as of unpredictability
(noise) inherent in such a coarse grained description.
\par
Our second motivation is the necessity of
having a technique reduing the description of a field system at
 the particle level. This is important in the context of studying open
field-theoretic quantum systems. We can derive master equations
describing the evolution of reduced density matrix of a field system,
by tracing out over the states of the environment. The environment can
for instance be a thermal bath or another field. We are then naturally
led to the question of what the evolution according to the master
equation will imply for the particles corresponding to the field. Our
preoccupation is particularly with the case of QED, where we can
obtain the evolution equation for the spinor field by tracing out the
states of the photon field \cite{AnZo}. The translation of this into 
a particle description is related to the notion of environment induced
decoherence \cite{BaCa} and the possibility of dynamicalorigin of the
charge superselection rules \cite{GKZ}.
\par
In this letter, we are primarily interested in the mathematical
aspects of the reduction of the field description into the particle
level. We are therefore going to consider the simple case of a $g
\phi^3$ scalar field theory, even though we do not expect to obtain
results of any physical interest. We should stress though, that the
techniqua developed here is easily generalized to apply for any field
theory that can be treated perturbatively and for all (even
non-unitary) dynamics.
 
\section{Preliminaries}
\paragraph{The Zwanzig technique}
For the construction of the master equation  we employ the Zwanzig projection
technique ( for review and discussion see \cite{Zeh,LiWe, RaMi}). 
The main concept in this formalism is the representation of the coarse
graining operation by an indempotent mapping {\bf P} in the space of
states
\begin{equation}
\rho \rightarrow \rho_{rel} =  {\bf P} \rho
\end{equation}
The irrelevant part of the state is then given by 
\begin{equation}
\rho_{irr} = ({\bf 1} - {\bf P}) \rho
\end{equation}
Considering unitary dynamics for the full theory 
\begin{equation}
i \frac{\partial \rho}{\partial t} = {\bf L} \rho \equiv [H,\rho]
\end{equation}  
we can show \cite{Zeh} that the evolution equation for the relevant part is given by
\begin{eqnarray}
i \frac{\partial \rho_{rel}(t)}{ \partial t} = {\bf PL} \rho_{rel}(t)
+ {\bf PL}  e^{-i ({\bf 1} - {\bf P}) {\bf L}t} \rho_{irr}(0) -i
\int_0^t d \tau {\bf G}(\tau) \rho_{rel}(t-\tau)
\end{eqnarray}
Here ${\bf G}$ stands for the kernel 
\begin{equation}
{\bf G}(\tau) = {\bf PL}({\bf 1} - {\bf P}) e^{-i  {\bf L} \tau} ({\bf 1} -
{\bf P}) {\bf LP}\end{equation}
In equation (3.) the third term corresponds to non-local in time noise
and dissipation while the second is essentially a time dependent
``external force '' acting on the relevant part of the state.
\par
In order to apply the Zwanzig method to our problem we need consider
coarse-graining operations of the form
\begin{equation}
\rho_{rel} = {\bf P} \rho = P_n \rho P_n
\end{equation}
where $P_n$ is the projection operator corresponding to the n-particle
subspace. The form of these operators is particularly transparent in
the Bargmann representation, in which we are going to carry our
calculations.  We here make a small digression for purposes of
establishing notation and assembling a number of useful formulas
\cite{Ber}. 

\paragraph{The Bargmann representation}
In the Bargmann representation one can associate to each operator
$\hat{A}$ on ${\cal F}$ a functional of $\alpha({\bf x})$ $\alpha^*({\bf x})$
where $\alpha({\bf x})$,$\alpha^*({\bf x})$ are elements of $H$ (complex valued
functions on a Cauchy  hypersurface $Sigma$ of Minkowski
spacetime). We actually have two choices for these functionals
 \begin{eqnarray}
\tilde{A}(\alpha^*,\alpha) = \sum \frac{1}{(n!m!)^{1/2}} \int
A_{mn}({\bf x}_1, \ldots, {\bf x}_m \mid {\bf y}_1, \ldots, {\bf y}_n) \nonumber \\
\alpha^*({\bf x}_1) \ldots \alpha^*({\bf x}_m) \alpha({\bf y}_n) \ldots \alpha({\bf y}_1) d^mx
d^ny
\end{eqnarray}
in terms of the distributions $A_{mn}$ of the matrix representation of
the operator, and
\begin{eqnarray}
A(\alpha^*,\alpha) = \sum  \int
A_{mn}({\bf x}_1, \ldots, {\bf x}_m \mid {\bf y}_1, \ldots, {\bf y}_n) \nonumber \\
\alpha^*({\bf x}_1) \ldots \alpha^*({\bf x}_m) \alpha({\bf y}_n) \ldots \alpha({\bf y}_1) d^mx
d^ny
\end{eqnarray}
in terms of the distributions $K_{mn}$ appearing in the normal
representation of the operator $\hat{A}$
\begin{eqnarray}
\hat{A} = \sum \int K_{mn}({\bf x}_1,\ldots,{\bf x}_n|{\bf y}_1,\dots, {\bf y}_m) \nonumber \\
\hat{a}^{ \dagger}({\bf x}_1) \ldots \hat{a}^{\dagger}({\bf x}_m) \hat{a}({\bf y}_1) \ldots \hat{a}({\bf y}_m)
d^mx d^ny
\end{eqnarray}
where $a({\bf x})$ and $a^{\dagger}({\bf x})$ satisfy the commutation relations 
\begin{equation}
[\hat{a}({\bf x}),\hat{a}^{\dagger} ({\bf x})] =  \delta ({\bf x}-{\bf x}')
\end{equation}
The two functionals are related by
\begin{equation}
A(\alpha^*,\alpha) = \tilde{A}(\alpha^*,\alpha) \exp \left[ - \int
\alpha^*({\bf x}) \alpha({\bf x}) dx \right]
\end{equation}
The functional corresponding to the product of operators
$\hat{C} = \hat{A} \hat{B}$ is given by
\begin{equation}
\tilde{C}(\alpha^*,\alpha) = \int D \beta^* D \beta \tilde{A}(\alpha^*,\beta)
\tilde{B}(\beta^*,\alpha) e^{- \beta^* \beta}
\end{equation}
In terms of the creation and annihilation operators $\hat{a}({\bf x})$ and
$\hat{a}^{\dagger}({\bf x})$ the Hamiltonian for a free field reads
\begin{equation}
\hat{H_0} = \frac{1}{2} \int dx dx' \hat{a}^{\dagger} ({\bf x}) h({\bf x},{\bf x}') \hat{a}({\bf x}')
\end{equation}
with 
\begin{equation}
h({\bf x},{\bf x}') = \int dk e^{-i{\bf k}({\bf x}-{\bf x}')} \omega_{\bf k}
\end{equation}
 Here $dk$ stands for the measure $d^3k / (2 \pi)^3$.
\par
The evolution operator $\hat{U}_0(t) =
e^{-it\hat{H_0}}$  reads in functional form
\begin{equation}
\tilde{U_0} (\alpha^*, \alpha;t) = \exp \left[ \int dx \alpha^*({\bf x})
\Delta({\bf x}-{\bf x}';t)
\alpha({\bf x}') \right]
\end{equation}
where 
\begin{equation}
\Delta({\bf x}-{\bf x}';t) = \int dk e^{-i  {\bf k}({\bf x}-{\bf x}')} e^{-i \omega_{\bf k} t}
\end{equation}
We also note the fundamental forula of Gaussian functional integration
\begin{eqnarray}
\int D \beta^* D \beta \beta^*({\bf x}_1) \ldots \beta^*({\bf x}_m) \beta({\bf x}'_1)
\ldots \beta({\bf x}'_n) e^{- \beta^* \beta + f^* \beta + \beta^* f}
\nonumber \\
= 
\frac{\delta^{m+n}} {\delta f({\bf x}_1) \ldots \delta f({\bf x}_m) \delta f({\bf x}'_1)
\ldots \delta f({\bf x}'_n)} e^{f^* f}
\end{eqnarray} 

\paragraph{Index notation}
 It is more convenient to represent the functional using an index
notation. To any function or distribution assign an abstract index
 to each of its arguments. The index is lower or upper  according to
whether the corresponding argument is
integrated out with an $\alpha$ or an $\alpha^*$ respectively. Hence
we represent
$\alpha({\bf x})$
with $\alpha^a$ and $\alpha^*({\bf x})$ with $\alpha^*_b$ and
$\Delta({\bf x}-{\bf x}';t)$ with $\Delta_{\; a}^b$. For example, the potential
operator to encounter in the following
\begin{equation}
\hat{V} = :\int dx \frac{g}{3!} \hat{\phi^3}:
\end{equation}
can be represented as
\begin{equation}
V(\alpha^*,\alpha) = \frac{g}{3!} \left( V_{abc} \alpha^a \alpha^b
\alpha^c + 3 \alpha^*_a V^a_{\; \; bc} \alpha^b \alpha^c + 3 \alpha^*_a
\alpha^*_b V^{ab}_{\quad c} \alpha^c + \alpha^*_a \alpha^*_b \alpha^*_c
V^{abc} \right)
\end{equation} 
with the correspondence
\begin{eqnarray}
V_{abc} & \leadsto \int \prod_{i=1}^{3} \frac{dk_i}{(2 \omega_{{\bf k}_i})^{1//2}}
e^{-i({\bf k}_1{\bf x}_1+{\bf k}_2 {\bf x}_2 +{\bf k}_3 {\bf x}_3)} (2 \pi)^3 \delta({\bf k}_1 + {\bf k}_2 + {\bf k}_3)
 \\
V^a_{ \; \; bc} & \leadsto \int \prod_{i=1}^{3} \frac{dk_i}{(2
\omega_{{\bf k}_i}^{1/2} )}
e^{-i(-{\bf k}_1{\bf x}_1+{\bf k}_2 {\bf x}_2 +{\bf k}_3 {\bf x}_3)} (2 \pi)^3 \delta({\bf k}_1 + {\bf k}_2 + {\bf k}_3)
 \\
V^{ab}_{\quad  c} & \leadsto \int \prod_{i=1}^{3} \frac{dk_i}{(2
\omega_{{\bf k}_i})^{1/2} }
e^{-i(-{\bf k}_1{\bf x}_1-{\bf k}_2 {\bf x}_2 +{\bf k}_3 {\bf x}_3)} (2 \pi)^3 \delta({\bf k}_1 + {\bf k}_2 + {\bf k}_3)
 \\
V_{abc} & \leadsto \int \prod_{i=1}^{3} \frac{dk_i}{(2 \omega_{{\bf k}_i})^{1/2}}
e^{i({\bf k}_1{\bf x}_1+{\bf k}_2 {\bf x}_2 +{\bf k}_3 {\bf x}_3)} (2 \pi)^3 \delta({\bf k}_1 + {\bf k}_2 + {\bf k}_3)
\end{eqnarray}
The inversion of the indices essentially amounts to the complex
conjugation of the distribution.

\section{ The master equation}
Now in the Bargmann representation the projection operator $\hat{P}_n$
corresponds to the functional
\begin{equation}
\tilde{P}_n( \alpha^*,\alpha) = \frac{1}{n!} (\alpha^* \alpha)^n
\end{equation}
Substituting this into the pre-master equation(3.8) we obtain
\begin{eqnarray}
i \frac{\partial \hat{\rho}(t)}{ \partial t} =
[\hat{H}_0,\hat{\rho}(t)] + [\hat{P}_n\hat{V} \hat{P}_n, \hat{\rho}(t)]
 \hspace{5cm} \nonumber \\
- i \int_0^t d \tau \left[ -\hat{A}_n(\tau) \hat{\rho}(t-\tau) \hat{B}_n ^{\dagger}(\tau) -
\hat{B}_n(\tau) \hat{\rho}(t-\tau) \hat{A}_n^{ \dagger} (\tau)   \right. \nonumber \\
\left.
+ \hat{C}_n(\tau) \rho(t-\tau) \hat{C}_n(-\tau)
+ \hat{C}^{ \dagger}_n (\tau) \hat{\rho}(t-\tau) \hat{C}^{\dagger}_n
(-\tau) \right] +
 \hat{F}_{res}(t)
\end{eqnarray}
where the operator-valued kernels $\hat{A}$, $\hat{B}$, $\hat{C}$ read
\begin{eqnarray}
\hat{A}_n(t) &=& \hat{P_n} \hat{V} (1-\hat{P_n})
e^{-i(\hat{H}_0+\hat{V})t} (1-\hat{P_n}) \hat{V} \hat{P_n} \\ 
\hat{B}_n(t) &=& \hat{P_n} e^{-i(\hat{H}_0 + \hat{V})t}\hat{P_n}                \\
\hat{C}_n(t) &=& \hat{P_n} \hat{V} (1-\hat{P_n})
e^{-i(\hat{H}_0+\hat{V})t} \hat{P_n}         
\end{eqnarray}
and the ``residual force'' operator $\hat{F}_{res}(t)$ 
corresponding to the effect of the
initial irrelevant part of the state is
\begin{eqnarray}
\hat{F}_{res}(t) &=& \hat{P}\hat{V}(1-\hat{P})
e^{-i(\hat{H}_0+\hat{V})t} \hat{\rho}_{irr}(0)
e^{i(\hat{H}_0+\hat{V})t}\hat{P} 
\nonumber \\ 
&-& \hat{P}e^{-i(\hat{H}_0+\hat{V})t} \hat{\rho}_{irr}(0)
e^{i(\hat{H}_0+\hat{V})t} (1-\hat{P}) \hat{V} \hat{P}  
\end{eqnarray}
For our choice of initial condition, that the system lies within the
n-particle subspace at $t = 0$, the residual force term vanishes and
our system of equations becomes autonomous.
\paragraph{ The lowest order in perturbation expansion}
The lowest order in the perturbation expansion is obtained by
replacing the operator $e^{-iHt}$ in our expressions for the kernel for
its free counterpart $e^{-iH_0t}$.
\par
Let us see what happens to the non-local terms when making this
approximation. The operator $\hat{C}_n$ vanishes since $P$ commutes with
$e^{-iH_0 \tau}$ and hence can be made to act on $1-P$. Hence the
non-local term in the master equation can be written
\begin{eqnarray}
-i \int^t_0 d \tau \left( -P_nV(1-P_n) (e^{-iH_0 \tau} V e^{iH_0 \tau})P_n 
 \rho(t-\tau) e^{i H_0 \tau} P_n \right. \nonumber \\
\left. - P_n e^{-iH_0 \tau} \rho(t-\tau) e^{iH_0 \tau} e^{-iH_0 \tau} V
 e^{iH_ \tau} (1-P)VP_n \right)
\end{eqnarray}
Now  $ || e^{-iH_0 \tau} \rho(t-\tau) e^{iH_0 \tau} - \rho(t) || =
O(g) $, hence by restricting ourselves to the second order in the
coupling constant we get a master equation that is local in time
\begin{equation}
i \frac{\partial \rho}{ \partial t} = [H_0,\rho] + [P_nVP_n,\rho] +i \left( L_n(t)
\rho + \rho L_n^{\dagger}(t) \right) + F_{res}(t)
\end{equation}
Here $L_n(t)$ stands for the operator:
\begin{equation}
\hat{L}_n(t) = P_nV(1-P_n) \left[ \int_0^t d \tau V(-\tau) \right] P_n
\end{equation}
where $V(\tau)$ stands for the time evolution of the operator $V$
according to the free Hamiltonian.
\par
We now set forth to calculate $\hat{L}$ for the particular cases of $n = 1$ and $ n =
2$. For this we need an expression for the operator $\hat{W}(t)  =
\int_0^t d \tau V(- \tau) $. The corresponding functional reads
 \begin{equation}
\tilde{W}(t)(\alpha^*,\alpha) = \frac{g}{3!} \left( W_{abc}
\alpha^a \alpha^b 
\alpha^c + 3 \alpha^*_a W^a_{\; \; bc} \alpha^b \alpha^c + 3 \alpha^*_a
\alpha^*_b W^{ab}_{\quad c} \alpha^c + \alpha^*_a \alpha^*_b \alpha^*_c
W^{abc} \right)
\end{equation} 
with 
\begin{eqnarray}
W_{abc}  \leadsto \int \prod_{i=1}^{3} \frac{dk_i}{(2 \omega_{k_i})^{1//2}}
e^{-i({\bf k}_1{\bf x}_1+{\bf k}_2 {\bf x}_2 +{\bf k}_3 {\bf x}_3)}  \hspace{5cm} \nonumber \\
\times \frac{e^{-i(\omega_{{\bf k}_1} +\omega_{{\bf k}_2}
+\omega_{{\bf k}_3})t} - 1}{-i(\omega_{{\bf k}_1} +\omega_{{\bf k}_2}
+\omega_{{\bf k}_3})} 
 (2 \pi)^3 \delta({\bf k}_1 + {\bf k}_2 + {\bf k}_3) \hspace{1cm}
 \\
W^a_{ \; \; bc}  \leadsto \int \prod_{i=1}^{3} \frac{dk_i}
{(2 \omega_{{\bf k}_i}^{1/2} )}
e^{-i(-{\bf k}_1{\bf x}_1+{\bf k}_2 {\bf x}_2 +{\bf k}_3 {\bf x}_3)}  \hspace{5cm} \nonumber \\
\times \frac{e^{-i(-\omega_{{\bf k}_1} +\omega_{{\bf k}_2}
+\omega_{{\bf k}_3})t} - 1}{-i(-\omega_{{\bf k}_1} +\omega_{{\bf k}_2}
+\omega_{{\bf k}_3})} 
 (2 \pi)^3 \delta({\bf k}_1 + {\bf k}_2 + {\bf k}_3) \hspace{1cm}
 \\
W^{ab}_{\quad  c}  \leadsto \int \prod_{i=1}^{3} \frac{dk_i}{(2
\omega_{{\bf k}_i})^{1/2} }
e^{-i(-{\bf k}_1{\bf x}_1-{\bf k}_2 {\bf x}_2 +{\bf k}_3 {\bf x}_3)} \hspace{5cm} \nonumber \\
\times \frac{e^{-i(-\omega_{{\bf k}_1} -\omega_{{\bf k}_2}
+\omega_{{\bf k}_3})t} - 1}{-i(-\omega_{{\bf k}_1} +-\omega_{{\bf k}_2}
+\omega_{{\bf k}_3})} 
(2 \pi)^3 \delta({\bf k}_1 + {\bf k}_2 + {\bf k}_3)  \hspace{1cm}
 \\
W^{abc}  \leadsto \int \prod_{i=1}^{3} \frac{dk_i}{(2 \omega_{{\bf k}_i})^{1/2}}
e^{i({\bf k}_1{\bf x}_1+{\bf k}_2 {\bf x}_2 +{\bf k}_3 {\bf x}_3)}  \hspace{5cm} \nonumber \\
\times \frac{e^{i(\omega_{{\bf k}_1} +\omega_{{\bf k}_2}
+\omega_{{\bf k}_3})t} - 1}{i(\omega_{{\bf k}_1} +\omega_{{\bf k}_2}
+\omega_{{\bf k}_3})} 
(2 \pi)^3 \delta({\bf k}_1 + {\bf k}_2 + {\bf k}_3) \hspace{1cm}
\end{eqnarray}
We can easily calculate 
\begin{eqnarray}
\widetilde{P_1V}(\alpha^*,\alpha) &=& \frac{g}{3!} \left( (\alpha^* \alpha)
V_{abc}\alpha^a \alpha^b \alpha^c + 3 \alpha^*_a
 V^a_{\; \; bc}
\alpha^b \alpha^ c \right) \nonumber \\
\widetilde{P_2V}(\alpha^*,\alpha) &=& \frac{g}{3!} \left( \frac{1}{2} 
(\alpha^* \alpha)^2 + 
V_{abc}\alpha^a \alpha^b \alpha^c \right. \nonumber \\
&\,& \hspace{1cm} \left. + 3 (\alpha^* \alpha) \alpha^*_a
V^a_{\; \; bc} 
\alpha^b \alpha^ c + 3 \alpha^*_a
\alpha^*_b V^{ab}_{\quad c} \alpha^c \right)
\end{eqnarray}
and from that with repeated application of equation (2.17) we get
\begin{equation}
\tilde{L_1}(t)(\alpha^*,\alpha) = \frac{g^2}{2}\left( \frac{1}{3!} V_{abc}
W^{abc} + \alpha^*_a (V^a_{\;\; cd} W^{cd}_{\;\;\;b} + V_{bcd}W^{cda})
\alpha^b \right)
\end{equation}
 and
\begin{eqnarray}
\tilde{L_2}(t)(\alpha^*,\alpha) = \frac{g^2}{4} \left( \frac{1}{3}
(\alpha^* \alpha)^2 V_{abc}W^{abc} \right. \nonumber \\
\left. + (\alpha^* \alpha)\alpha^*_a (2 V^a_{\;\;cd} W^{cd}_{\;\;\;b}
+ V_{bcd} W^{cda}) \alpha^b \right. \nonumber \\
\left. +2 \; \alpha^*_a \alpha^*_b (2 V^b_{\;\;ce} W^{ea}_{\;\;\; d} +
V_{cde}W^{abe}) \alpha^c \alpha^d \right) 
\end{eqnarray}

\paragraph{Renormalization}
Since we performed a perturbation expansion, it is inavoidable to
encounter divergences. We readily see that the term $Z =
V_{abc}W^{abc}$ as a contraction of two distribution has a
$\delta$-function divergence
\begin{equation}
Z (t) = \int \prod_{i=1}^{3} \frac{dk_i}{(2 \omega_{{\bf k}_i}}
\frac{e^{-i(\omega_{{\bf k}_1} +\omega_{{\bf k}_2}
+\omega_{{\bf k}_3})t} - 1}{-i(\omega_{{\bf k}_1} +\omega_{{\bf k}_2}
+\omega_{{\bf k}_3})} 
(2 \pi)^6 \delta({\bf k}_1 + {\bf k}_2 + {\bf k}_3) \delta(0)
\end{equation}
These terms can be removed from the expression of the operator
$\hat{L}(t)$ by a renormalization of the density matrix. To see this
it is sufficient to remark that the divergent part of $\hat{L}$ is
proportional to the projection operator $P_n$. Hence in the master
equation these divergencies are contained in the terms 
\begin{equation}
i \frac{g^2}{24} \left( Z(t) + Z^*(t) \right) \rho(t)
\end{equation}
These terms can be absorbed through a time dependent
renormalization of the density matrix (essentially wave function
renormalization \cite{Ber})
 \begin{equation}
\rho(t) \rightarrow \rho_{ren}(t) = \exp \left( \frac{g^2}{24} \int_0^t d \tau [Z(\tau) +
Z^*(\tau)] \right) \rho(t)
\end{equation}
which clearly presents positivity, since the term in the exponent is
real.
\par
This is not the only term giving rise to infinities in the master
equation, but it is more convenient to deal with the others when
bringing the master equation into a suitable form.
\par
Having removed these terms we get the following expressions for the operators
\begin{eqnarray} 
\tilde{L}_1(t) = \frac{g^2}{2} \alpha^*_a F^a_{\;\;b} \alpha^b
\nonumber \\
\tilde{L}_2(t) = \frac{g^2}{4} \left (\alpha^* \alpha) \alpha^*_a
F^a_{\;\;b} \alpha^b + \alpha^*_a \alpha^*_b G^{ab}_{\;\;\;cd}
\alpha^c \alpha^d \right)
\end{eqnarray}
with
\begin{eqnarray}
 F^a_b \leadsto \prod_{i=1}^{3} \frac{dk_i}{(2 \omega_{{\bf k}_i})}
e^{-i{\bf k}_1({\bf x}-{\bf x}')} 
 (2 \pi)^3 \delta({\bf k}_1 + {\bf k}_2 + {\bf k}_3) \hspace{3cm}
\\
\nonumber \\ 
\times \left( \frac{e^{-i(\omega_{{\bf k}_1} - \omega_{{\bf k}_2}
-\omega_{{\bf k}_3})t} - 1}{-i( \omega_{{\bf k}_1} -\omega_{{\bf k}_2}
- \omega_{{\bf k}_3})} + \frac{e^{i(\omega_{{\bf k}_1} + \omega_{{\bf k}_2}
+ \omega_{{\bf k}_3})t} - 1}{i(\omega_{{\bf k}_1} +\omega_{{\bf k}_2}
+ \omega_{{\bf k}_3})}  \right) \hspace{2cm}
\\ 
 G^{ab}_{\; \; \; cd} \leadsto  \int \prod_{i=1}^{2} 
\frac{dk_i}{(2 \omega_{{\bf k}_i})^{1/2}}
\frac{dk'_i}{(2 \omega_{{\bf k}'_i})^{1/2}} \frac{dk_3}{2 \omega_{{\bf k}_3}} \hspace{3cm}
\nonumber \\
 \times (2 \pi)^6 \delta({\bf k}_1+{\bf k}_2+{\bf k}_3) \delta({\bf k}'_1 + {\bf k}'_2
-{\bf k}_1 -{\bf k}_2) e^{i({\bf k}_1{\bf x} +{\bf k}_2 {\bf x}' -{\bf k}'_1 y - {\bf k}'_2 y')} \hspace{1cm} 
\nonumber \\
\times 
\left(  \frac{e^{-i(\omega_{{\bf k}'_1} + \omega_{{\bf k}'_2}
+ \omega_{{\bf k}_3})t} - 1}{-i(\omega_{{\bf k}'_1} +\omega_{{\bf k}'_2}
+ \omega_{{\bf k}_3})} + 2 \frac{e^{-i(-\omega'_{{\bf k}_1} - \omega'_{{\bf k}_2}
+ \omega_{{\bf k}_3})t} - 1}{-i(- \omega'_{{\bf k}_1} -\omega'_{{\bf k}_2}
+ \omega_{{\bf k}_3})}  \right) \hspace{2cm}
\end{eqnarray}

\section{One particle}
 Let us now concentrate to the case of the one particle, and in
particular try to establish the form to which the master equation
(3.8) reduces in the non-relativistic regime.
\par
When restricting to the one-particle subspace the reduced density
matrix is of the form
\begin{equation}
\tilde{\rho} = \alpha^*_a \rho^a_{\;\;b} \alpha^a
\end{equation}
with $\rho^a_{\;\;b}$ essentially corresponding the standard form of
the density matrix $\rho({\bf x},{\bf x}')$ in the position representation.
Substituting (4.1),(3.) and (3.) in the master equation (3.18) we
obtain the evolution equation
\begin{equation}
i \frac{\partial}{\partial t} \rho^a_{\;\;b} = h^a_{\;\;c}
\rho^c_{\;\;b} - \rho^c_{\;\;b} h^a_{\;\;c} + i \frac{g^2}{2} \left(
F^a_{\;\;c} \rho^c_{\;\;b} + \rho^c_{\;\;b} F^a_{\;\;c} \right)
\end{equation}
It is more convenient to consider the density matrix in the momentum
representation
\begin{equation}
\rho({\bf k},{\bf k}') = \int dx dx' e^{-i{\bf k}{\bf x} +i{\bf k}'{\bf x}'} \rho({\bf x},{\bf x}')
\end{equation} 
in which we obtain the transparent form
\begin{equation}
i \frac{\partial}{\partial t} \rho({\bf k},{\bf k}') =   ( \omega_{\bf k} - \omega_{{\bf k}'})
\rho({\bf k},{\bf k}') +i g^2 \left[ F({\bf k},t) + F^*({\bf k}',t) \right]
\rho({\bf k},{\bf k}')
\end{equation}
where the function $F$ is given by
\begin{eqnarray}
F({\bf k},t) = \frac{i}{8 \omega_{\bf k}} \int \frac{d k'}{\omega_{{\bf k}'}
\omega_{{\bf k}+{\bf k}'}} \nonumber \\ \times
\frac{e^{-i \omega_{\bf k} t} 
\left( \omega_{\bf k} \cos(\omega_{{\bf k}'}
+ \omega_{{\bf k}+{\bf k}'})t + i (\omega_{{\bf k}'} + \omega_{{\bf k}+{\bf k}'}) \sin(\omega_{{\bf k}'}
+ \omega_{{\bf k}+{\bf k}'})t \right) - \omega_{\bf k} }{\omega_{\bf k}^2 - (\omega_{{\bf k}'}
+ \omega_{{\bf k}+{\bf k}'})^2}
\end{eqnarray}
The function $F$ is actually divergent, but the divergence
(logarithmic) is
essentially contained in its ${\bf k} = 0$ argument. Therefore, we can
define the renormalized function $F_{ren}({\bf k},t) = F({\bf k},t) - F(0,t)$
and absorb the divergent part $(F(0,t) + F^*(0,t) = Z'(t)$ in a
renormalization of the density matrix
\begin{equation}
\rho(t) \rightarrow \rho_{ren}(t) = \exp \left( g^2 \int^t_0 Z'(\tau) \; d
\tau \right) \rho(t)
\end{equation}
\paragraph{The non-relativistic limit}
To obtain the non-relativistic limit it is sufficient to obtain the
lowest order term in the expansion of $F_{ren}({\bf k},t)$ in powers of
${\\bf {\bf k}}$ that is, take the approximation
\begin{equation}
F_{ren}({\bf k},t) \approx \frac{\partial^2 F}{\partial k^i \partial
k^j}|_{{\bf k}=0} k^i k^j  
\end{equation}
since the first order derivative of $F({\bf k},t)$ can easily be shown to vanish
due to symmetry arguments. Writing 
\begin{equation}
\frac{\partial^2 F}{\partial k^i \partial
k^j}|_{{\bf k}=0} (t) = \gamma(t) \delta^{ij} + i \zeta(t) \delta^{ij}
\end{equation}
as we expect again from symmetry requirements, we obtain the master
equation
\begin{equation}
i \frac{\partial}{\partial t} \hat{\rho} = (1 + 2 i m g^2\zeta(t))[
\hat{H}_0 ), \hat {\rho}] + i 2m g^2 \{ \hat{H}_0, \hat{\rho} \}
\end{equation}
\paragraph{Some comments}
The form of equation (4.9) is clearly unusual , but the interpretation is
not difficult. Essentially, the effect of the environment is to
complexify  the Hamiltonian, a fact which relates to  the
non-conservation of particles (it is easy to check that this is the
term that fails to preserve the trace). The term in the anticommutator
is a term exhibiting dissipation through coupling to energy. 
\par
Also the fact that we have considered a three-field interaction vertex
leads to the non-appearance of a potential term in the master
equation. This will be expected to appear when considering at least
the $n =3$ level. To obtain the non-relativistic potential one has to
consider theories with two relevant field lines at the vertex, as is
the cas of QED.

\end{document}